\documentclass[letterpaper, 10 pt, conference]{ieeeconf}  % Comment this line out if you need a4paper
\usepackage{multicol}
\usepackage[bookmarks=true]{hyperref}
\usepackage{float}
\usepackage{graphicx}
\usepackage{epstopdf}
\usepackage{booktabs}
\usepackage{tabularx}
\usepackage{multirow}
\usepackage{caption}
\usepackage{subcaption}
\usepackage{amssymb}
\usepackage{color}
\usepackage{amsmath}
\usepackage{multirow}
\usepackage{comment}
\usepackage{balance}
\usepackage{soul}
\usepackage[ruled, linesnumbered]{algorithm2e} %ruled,linenumbered

\usepackage[bottom]{footmisc}
\usepackage{tabularx}
\IEEEoverridecommandlockouts                              % This command is only needed if 
\usepackage{xcolor}
\newcommand{\mypara}[1]{\vspace{0.7em}\noindent\textbf{#1}}

\title{\LARGE \bf
	TactileSGNet: A Spiking Graph Neural Network for Event-based Tactile Object Recognition
}

\author{Fuqiang Gu, Weicong Sng, Tasbolat Taunyazov, and Harold Soh\\
	Dept. of Computer Science, School of Computing,\\
	National University of Singapore\\
	{\tt\small \{gufq, sngweicong, tasbolat, harold\}@comp.nus.edu.sg}
}

\begin{document}

	\maketitle
	\thispagestyle{empty}
	\pagestyle{empty}

	%%%%%%%%%%%%%%%%%%%%%%%%%%%%%%%%%%%%%%%%%%%%%%%%%%%%%%%%%%%%%%%%%%%%%%%%%%%%%%%%
	\begin{abstract}
		Tactile perception is crucial for a variety of robot tasks including  grasping and in-hand manipulation. New advances in flexible, event-driven, electronic skins may soon endow robots with touch perception capabilities similar to humans. These electronic skins respond asynchronously to changes (e.g., in pressure, temperature), and can be laid out irregularly on the robot's body or end-effector. However, these unique features may render current deep learning approaches such as convolutional feature extractors unsuitable for tactile learning. In this paper, we propose a novel spiking graph neural network for event-based tactile object recognition. To make use of local connectivity of taxels, we present several methods for organizing the tactile data in a graph structure. Based on the constructed graphs, we develop a spiking graph convolutional network. The event-driven nature of spiking neural network makes it arguably more suitable for processing the event-based data.  Experimental results on two tactile datasets show that the proposed method outperforms other state-of-the-art spiking methods, achieving high accuracies of approximately 90\% when classifying a variety of different household objects.
	\end{abstract}

	%%%%%%%%%%%%%%%%%%%%%%%%%%%%%%%%%%%%%%%%%%%%%%%%%%%%%%%%%%%%%%%%%%%%%%%%%%%%%%%%
	\section{Introduction}
	Object recognition is a basic perceptual skill that underlies many tasks, from driving a car to preparing a meal. Advances in machine vision have provided robots with excellent visual object recognition capabilities (e.g., \cite{gevers1999color, eitel2015multimodal}). But while vision serves as an important visual modality, it can fail to distinguish objects with similar visual features or in less-favorable conditions, e.g., under low-lighting or occlusion. In such cases, tactile sensing can provide important information (e.g., texture, roughness, friction), which has been applied in a variety of tasks including object recognition~\cite{kappassove2015tactile,navarro2012haptic,soh2012online}, slip detection~\cite{calandra2018more}, and texture recognition~\cite{taunyazov2019towards}.
	
	This study focuses on the challenging task of touch-based object recognition with \emph{event-driven} \emph{tactile} sensors~\cite{taunyazov20event,lee2019neuro}. Prior works (e.g., ~\cite{taunyazov2019towards, navarro2012haptic, kappassove2015tactile}) have mainly used standard synchronous tactile sensors with conventional machine learning approaches (e.g., convolutional neural networks~\cite{krizhevsky2012imagenet}). However, event-driven sensors are inherently different, both in terms of operation and data provided. Similar to event-based cameras~\cite{gehrig2019end, mitrokhin2019ev}, event tactile sensors asynchronously report changes in the environment and thus, provide event-based ``spikes'' where each taxel fires independently of the rest. Compared to standard synchronous frame-based sensors, event-driven sensing can achieve higher power-efficiency, better scalability, and lower latency. However, learning with these sensors remains in its infancy~\cite{Pfeiffer2018}.
	
	In this paper, we present TactileSGNet, a novel spiking graph neural network for object recognition using event-based tactile data. In contrast to convolutional neural networks for grid-structured real-valued data, our model operates on \emph{graph-structured spiking data}. This provides two key advantages: first, the model can better exploit local taxel structure that can be highly irregular, e.g., with  biologically-inspired configurations or flexible sensors that are wrapped around end-effectors. Second, spiking neural networks (SNNs) are also event-driven and can directly process the spike-based data provided by the sensors; this bypasses potentially expensive transformations from discrete events to real-valued frames. In addition, SNNs can be run on power-efficient neuromorphic processors such as the IBM TrueNorth~\cite{merolla2014million} and Intel Loihi~\cite{davies2018loihi}. 
	
	To our knowledge, TactileSGNet is the first event-driven graph neural network for tactile data. A related model is the recently proposed TactileGCN~\cite{garcia2019tactilegcn}, which uses a graph convolutional network (GCN)~\cite{kipf2016semi} for tactile object recognition. The key differences in this work is that TactileSGNet is event-driven (with spiking neurons) and we utilize a topology adaptive graph convolutional network (TAGCN)~\cite{du2017topology}; the TAGCN has been previously shown to achieve superior performance, whilst being computationally more efficient compared to standard GCNs. Indeed, our computational experiments on two existing event-based tactile datasets using the NeuTouch sensor~\cite{taunyazov20event} show that leveraging the TAGCN with spiking neurons achieves superior performance to other popular architectures. We also experiment with alternative approaches for constructing tactile graphs; results suggest automated methods, specifically nearest-neighbor and minimum spanning tree techniques, can achieve even better performance.  
	
	\section{Background \& Related Work}
	\label{sec:relatedwork}
	Our work combines the recent advances in graph neural networks (GNNs) and spiking neural networks (SNNs) for event-based tactile object recognition.
	In the following, we provide a brief overview of background and related work in these areas. Note that these research areas are broad and, due to space constraints, we cover representative work and refer readers wanting more details to more comprehensive survey articles. 
	
	\mypara{Tactile Sensing.}
	Tactile sensing provides a modality of information (e.g., roughness, textures, temperature) that is different from visual sensing; incorporating a sense of touch enables robots to better perceive physical environments. Tactile perception has been in many robot tasks such as object recognition~\cite{kappassove2015tactile,navarro2012haptic,soh2012online}, slip detection~\cite{calandra2018more}, and texture recognition~\cite{taunyazov2019towards}. 
	
	To date, several types of tactile sensors have been developed (see \cite{liu2017recent} for a survey); popular sensors include the BioTac\footnote{\url{https://www.syntouchinc.com/technology/}}, PPS\footnote{\url{https://pressureprofile.com/}}, and Tekscan\footnote{\url{https://www.tekscan.com/}}.
	In this paper, we focus on using NeuTouch, an event-based tactile sensor that have been proposed in recent work~\cite{taunyazov20event}. Little prior work exists for learning with event-based tactile data. Very recent work~\cite{taunyazov20event} proposed a multi-modal spiking network based on SLAYER~\cite{shrestha2018slayer}. Our work is different in that we explore graph spiking neural networks (rather than fully connected layers) with LIF (leaky integrate-and-fire)~\cite{wu2018spatio} neurons rather than SRM (spike response model) neurons~\cite{gerstner1995time}. 
	
	\begin{figure}
		\centering 
		\includegraphics[width=0.8\linewidth]{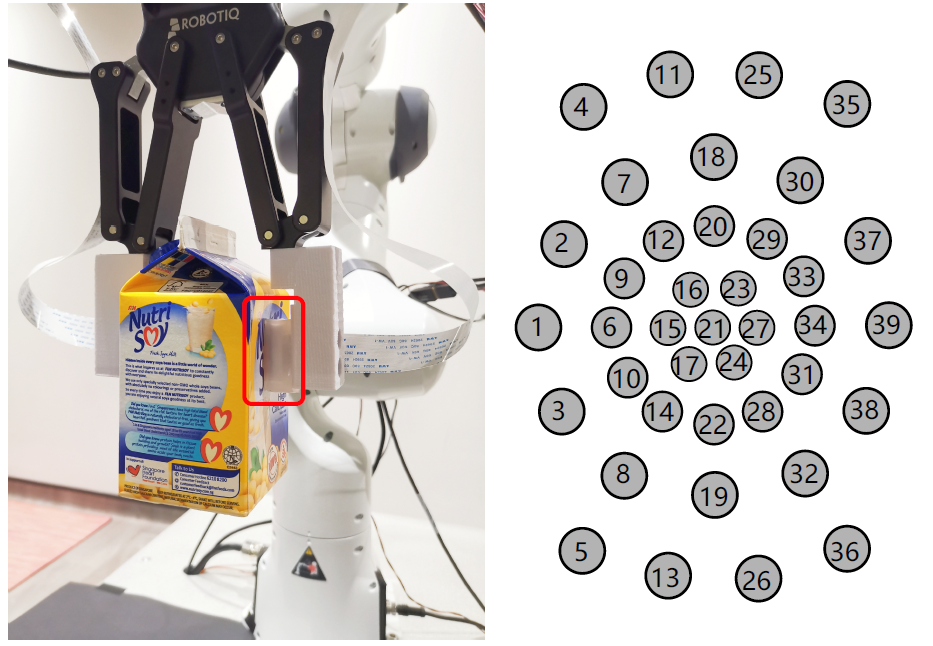}
		\caption{The NeuTouch sensor mounted on a Robotiq gripper and spatial distribution of 39 NeuTouch taxels.} 
		\label{fig:taxel_distribution} 
	\end{figure}
	
	\mypara{Graph Neural Networks (GNNs)} are a class of models that combine deep learning models and methods for structured data~\cite{battaglia2018relational}. GNNs have gained popularity of late due to their applicability in many fields, from social network mining to embedding logic into deep networks~\cite{yaqi2019embedding}.
	
	Of particular interest for this work are GCNs where the convolution operation is conducted in the spectral domain via trainable graph filters~\cite{bruna2013spectral,bianchi2019graph}. To reduce the computational cost of decomposition and projection in the frequency domain, graph filters are usually approximated using finite order polynomials. For example, in \cite{bruna2013spectral, defferrard2016convolutional}, the graph filters are approximated using high-degree Chebyshev polynomials of the graph Laplacian matrix. A popular GCN~\cite{kipf2016semi} approximates graph filters with the first-order Chebyshev polynomials of graph Laplacian. 
	More recent work~\cite{du2017topology} has proposed limiting the polynomials of the adjacency matrix (to maximum degree of two) to further reduce the complexity. In this study, we use the TAGCN to perform convolution on the tactile data due to its computational efficiency and demonstrated performance~\cite{du2017topology}.
	
	\mypara{Spiking Neural Networks (SNNs)}
	form a core approach in neuromorphic computing~\cite{roy2019towards}. SNNs are more biologically plausible than deep neural networks (DNNs), and can be executed on power-efficient neuromorphic hardware (e.g., the Intel Loihi~\cite{davies2018loihi}). SNNs can have the similar network topology as DNNs, but use different neuron models. Commonly-used neuron models for SNNs include the LIF~\cite{wu2018spatio} and SRM~\cite{gerstner1995time}. One issue in SNNs is that the spike function is non-differentiable, making it impossible to use backpropagation to train the network. To address this issue, several solutions have been proposed, such as converting DNNs to SNNs~\cite{esser2015backpropagation}, and approximating the derivative of the spike function~\cite{wu2019direct, shrestha2018slayer}. In this work, we use SNNs as they are able to directly handle spiking sensor data. 
 
	\section{Proposed Method: Learning With\\Tactile Graphs}
	\label{sec:method}
	
	In this section, we provide a description of our graph-based approach for learning from event-based tactile data. As previously mentioned, unlike visual \emph{pixels}, the \emph{taxels} for touch sensing may be structured in an irregular fashion.  Indeed, human touch sensors are distributed unevenly across the body (with correspondingly different neurological demands as illustrated by the popular cortical homunculus). 
	
	As artificial e-skins continue to develop in both capabilities and affordability, we anticipate that robots will incorporate flexible skins that provide similar (or possibly superior) touch sensing capabilities to humans. The tactile sensors may be ``wrapped'' around existing body parts or have taxels organized in irregular configurations. Consider the NeuTouch~\cite{taunyazov20event} used in our experiments; the NeuTouch is a biologically-inspired fingertip tactile sensor with 39 taxels that are arranged spatially in radial fashion (Fig. \ref{fig:taxel_distribution}). In the following, we will use the NeuTouch as our running example to describe our method, but note that our approach can be utilized with other sensors with different taxel configurations and layouts.

	\subsection{Tactile Graph Construction}
	\label{sec:tactile_graph}
	To process the data from tactile sensors, one could adopt standard convolutional layers used in deep neural networks~\cite{zapata2018non}. However, this would require ``forcing'' the data into a grid structure, which entails specifying an arbitrary grid size with zero-filled (or interpolated) cell values. Here, we undertake a more \emph{natural} approach by constructing a tactile graph based on the local spatial arrangement of the underlying taxels.
	
\begin{figure}
		\centering
		\begin{subfigure}[b]{0.3 \linewidth}
			\centering
			\includegraphics[width=1 in]{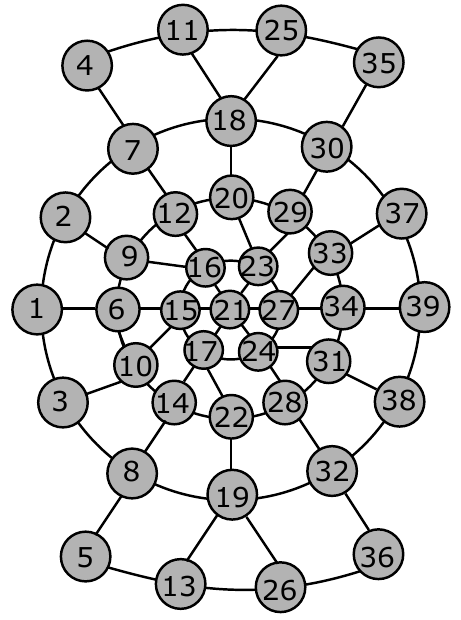} 
			\caption{Manual}
		\end{subfigure}
		~
		\begin{subfigure}[b]{0.3 \linewidth}
			\centering
			\includegraphics[width=1 in]{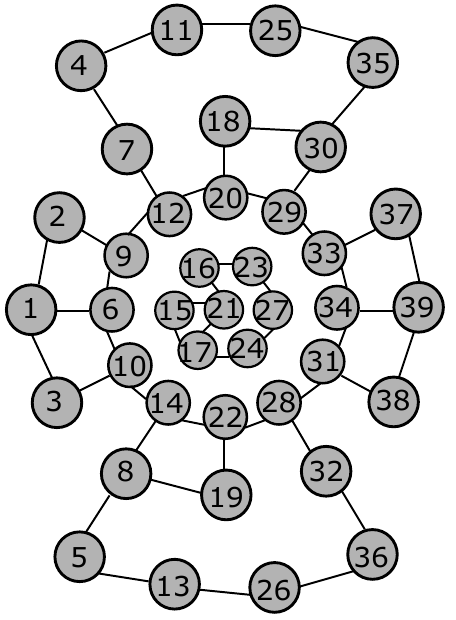} 
			\caption{kNN (k=2)}
		\end{subfigure}
		~
		\begin{subfigure}[b]{0.3 \linewidth}
			\centering
			\includegraphics[width=1 in]{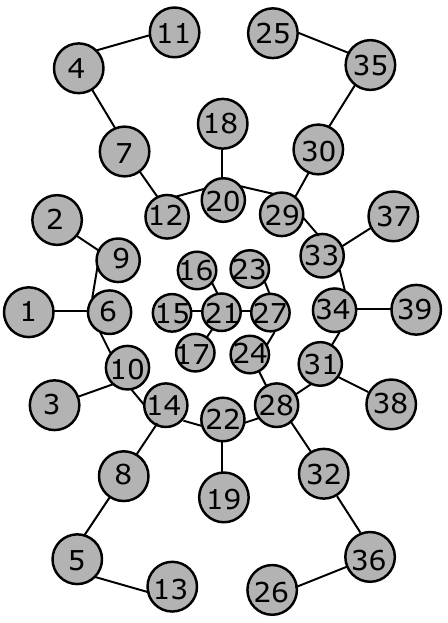} 
			\caption{MST ($\sigma_d=0$)}
		\end{subfigure}
		\caption{Tactile graphs constructed by different distance-based methods. (a) Graph constructed manually by-hand. (b) Graph obtained by kNN (k=2). (c) Graph generated using Kruskal's MST algorithm.}
		\label{fig:tactile_graph} 
	\end{figure}

	Let $G=(V, E)$ be a tactile graph, where $V$ is a set of $N$ nodes, and $E$ is a set of undirected edges\footnote{We focus on undirected edges, but our method also accommodates directed edges.}. The nodes are naturally mapped to taxels, but the edges have to be specified. We propose to leverage the spatial/geometric configuration of the points and introduce edges based upon the Euclidean distances between nodes $d(v_i,v_j) = \| v_i - v_j\|_2$. Here, we explore three different distance-based methods:
	\begin{enumerate}
		\item \textbf{Manual}, where edges are manually connected according to their physical proximity;
		\item \textbf{k-Nearest Neighbors (kNN)}, where each node is connected to its $k$ closest neighbors;
		\item \textbf{Minimum Spanning Tree (MST)}, where edges in a MST are added to the edge set of the graph, along with extra edges between any two nodes with distances smaller than a user-specified distance threshold $\sigma_d$.
	\end{enumerate}
	As a concrete example, Fig. \ref{fig:tactile_graph} illustrates the tactile graphs constructed by using the above methods for the NeuTouch. Our experiments will largely compare methods using the manual approach, but we include additional experiments that show how the graph connectivity affects performance on the object recognition task. 
	
	\begin{figure*}
		\centering
		\includegraphics[width= 0.95\linewidth]{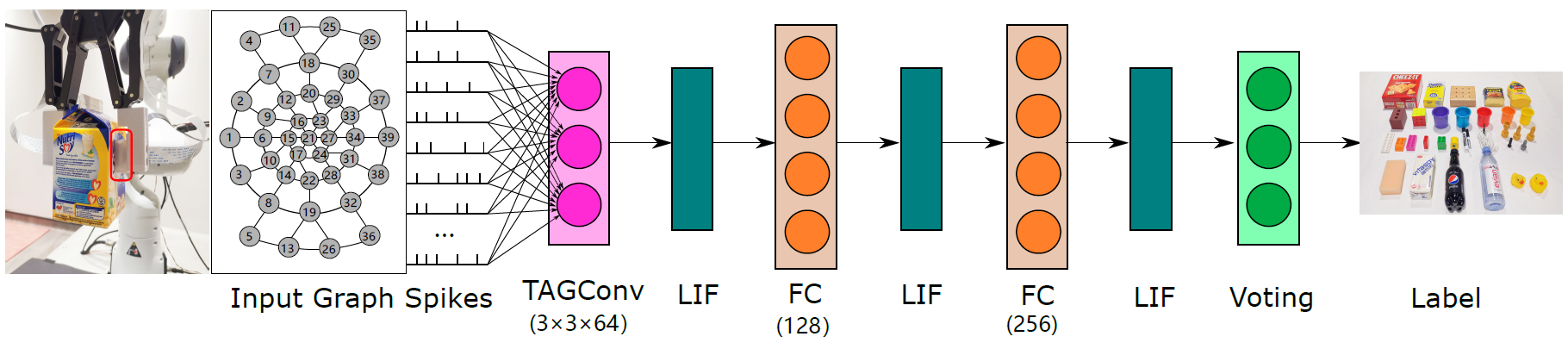}
		\caption{TactileSGNet Architecture. TactileSGNet is a spiking neural network (SNN) that processes input spikes from taxels with connectivity specified by an input graph. It comprises a graph convolutional layer, two fully-connected (FC) layers, and a voting layer.} 
		\label{fig:architecture} 
	\end{figure*}
	
	\begin{figure}
		\centering
		\includegraphics[width= 0.6 \linewidth]{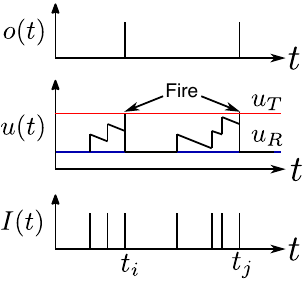} 
		\caption{ The dynamics of a LIF neuron. It takes as input binary spikes and outputs binary spikes. $I(t)$ represents the input signal to a neuron, $u(t)$ is the membrane potential of the neuron, and $o(t)$ is the output of the neuron. An output spike will be emitted from the neuron when its membrane potential surpasses the firing threshold $u_T$, after which the membrane potential will be reset to $u_R$.}  
		\label{fig:LIF_model} 
	\end{figure}
	
	\subsection{TactileSGNet}
	To process the data from our tactile graph, we propose a spiking neural network architecture we call TactileSGNet (shown in Fig. \ref{fig:architecture}). The network uses LIF neurons, and includes a topology adaptive graph convolutional  (TAGConv) layer~\cite{du2017topology}, fully-connected (FC) layers, and a final voting layer for classification. In the following, we describe each of these components:
	
	\mypara{LIF Activations.} In conventional convolutional neural networks, the most common activation functions are the ReLU \cite{glorot2011deep} and its variants (e.g., LReLU \cite{maas2013rectifier}). However, the ReLU activation function is unsuitable in SNNs. We use the LIF model, which is a popular model for describing the dynamics of spiking neurons~\cite{gerstner2002spiking, lee2016training, roy2019towards}. The dynamics of the LIF neuron is described by
	\begin{equation}
	\tau \frac{du(t)}{dt} = - u(t) + \sum_i w_i x_i,
	\label{eqn:LIF}
	\end{equation}
	where $u(t)$ represents the internal membrane potential of a neuron at time $t$, $\sum_i w_i x_i$ is the weighted summations of the inputs from pre-neurons, and $\tau$ is a time constant. Fig. \ref{fig:LIF_model} visualizes the dynamics of a LIF spiking neuron. 
	
	To better understand the membrane potential update, we can apply the Euler method to approximate the solution of the differential equation (\ref{eqn:LIF}) shown above~\cite{wu2019direct}. Then, the update of the membrane potential can be written as: 
	\begin{equation}
	u({t+1}) = (1 - \frac{dt}{\tau})u(t) + \frac{dt}{\tau} \sum_i w_i x_i,
	\end{equation}
	which can be further simplified as:
	\begin{equation}
	u({t+1}) = \beta u(t) + \sum_i w'_i x_i,
	\end{equation}
	where $\beta= 1 - \frac{dt}{\tau}$ can be considered as a decay factor, and $w'_i$ is the weight incorporating the scaling effect of $\frac{dt}{\tau}$.
	Thus, the LIF activation function $f_{LIF}$ is described as: 
	\begin{equation}
	f_{LIF} (u) =  \text{fire a spike}\ \text{\&} \ u(t) \gets u_R  \ , \text{if} \ u(t) \geq u_T
	\end{equation}
	where $u_R$ and $u_T$ are constants, representing the reset value and firing threshold, respectively. The LIF activation function indicates that a neuron will fire (i.e., output a spike) when its membrane potential reaches or surpasses a given threshold $u_T$. After the firing, its membrane potential will be reset to a value $u_R$. 
	
	\mypara{TAGConv Layer.} Compared to the popular graph convolution~\cite{kipf2016semi}, the TAG convolution~\cite{du2017topology} used in our network adapts to the topology of the input graph. A TAG convolution operation is defined as:
	\begin{equation}
	\mathbf{z}_f = \sum_{c=1}^{C} \mathbf{G}_{c,f} * \mathbf{x}_c + \mathbf{b}_f,
	\end{equation}
	where $\mathbf{z}_f$ is the $f$-th output feature map, $\mathbf{x}_c$ is the $c$-th input feature of all nodes ($\mathbf{x}_c \in \mathbb{R}^N$, where $N$ is the number of nodes), $C$ is the number of input features of each node. $\mathbf{b}_f$ is a learnable bias vector, $*$ is the convolution operator, and $\mathbf{G}_{c,f}$ is the $f$-th graph filter. To make the convolution operation work for arbitrary graph topologies, the graph filter needs to be carefully designed. One  approach is to define the graph filter with the normalized adjacency matrix of the graph, 
	\begin{equation}
	\mathbf{G}_{c,f} = \sum_{k=0}^{K} g_{c,f,k} \mathbf{A}^k, 
	\end{equation}
	where $g_{c,f,k}$ is the polynomial coefficient of the graph filter, and $\mathbf{A}$ is the normalized adjacency matrix. 
	
	\mypara{Fully-Connected (FC) Layer.} The FC layer is similar to those used in conventional neural networks, i.e., 
	\begin{equation}
	\mathbf{h} = \mathbf{W} \mathbf{x}  +  \mathbf{b},
	\end{equation}
	where $\mathbf{x}$ is the inputs from previous layer, $\mathbf{W}$ is the weight matrix, $\mathbf{b}$ is the bias vector, and $\mathbf{h}$ is the output feature. 
	
	\mypara{Voting Layer.} The voting layer is used to decode the network output. We adopt the voting strategy in \cite{diehl2015unsupervised, wu2019direct}; briefly, each output label is first associated to one of neurons in the voting layer. The predicted class is the one associated with the neuron with the largest number of votes (corresponding to spikes) averaged over the time window. 
	
	\subsection{Training }
	To train the network, we define the loss function that captures the mean squared error between the label vector $\mathbf{y}$ and the averaged voting results over a given time window, 
	\begin{equation}
	\mathcal{L} = \left\Vert \mathbf{y} - \frac{1}{T} \sum_{t=1}^T \mathbf{U} \mathbf{o}^t \right\Vert^2
	\end{equation}
	where $\mathbf{U}$ is the voting matrix, and $\mathbf{o}^t$ is the output feature from the last layer at time $t$.
	
	In non-spiking neural networks, one can train a network by minimizing the loss function via standard backpropagation. However, spikes are non-differentiable. Fortunately, we can approximate the derivative of the spike function, which has been shown to be effected on various tasks~\cite{lee2016training, zenke2018superspike, shrestha2018slayer}. In this study, we use the rectangular function $f(u)$ to approximate the derivative of the spike function due to its simplicity and reported performance~\cite{wu2019direct, wu2018spatio},  
	\begin{equation}
	f(u) = \frac{1}{a} \textrm{sign} \left(\left |u- u_T \right| < \frac{a}{2}\right)
	\end{equation}
	where $a$ is a width parameter. 
	
	\section{Experimental Results}
	The primary objective of our experiments was to evaluate different  architectures for event-based tactile object recognition. To this end, we compare alternative architectures including multi-layer perceptron (MLP) and convolutional neural networks (CNN) against our model. We also sought to understand the potential benefits of using the adaptive TAGConv layer, rather than a standard GCN. In the above experiments, we focused on the manually-designed input graph. We ran a second set of experiments to compare the three different graph construction methods described in Sec.~\ref{sec:tactile_graph}.
	
	\subsection{Datasets}
\begin{figure}
		\centering
		\begin{subfigure}[b]{0.45 \linewidth}
			\centering
			\includegraphics[height =1 in]{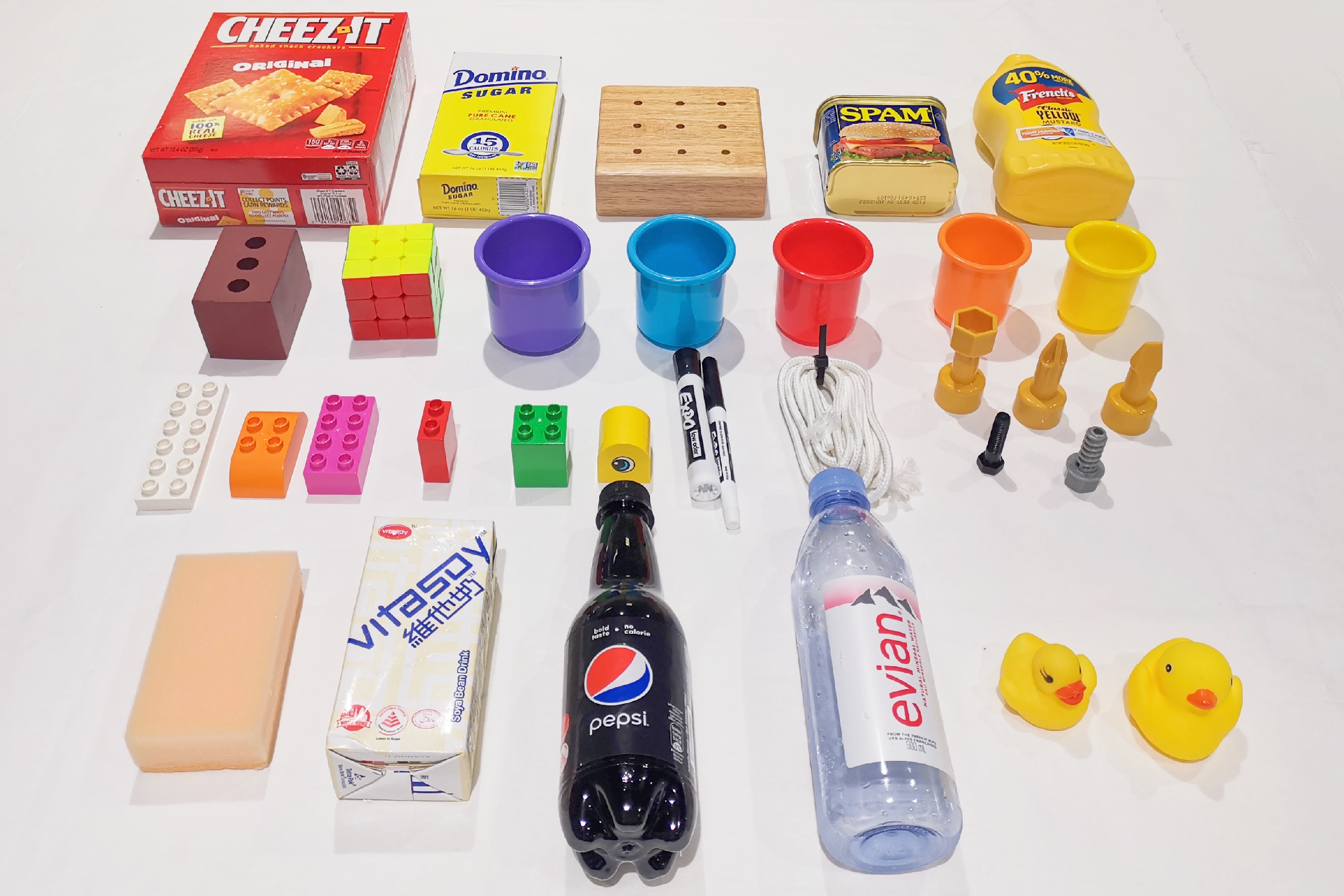} 
			\caption{EvTouch-Objects}
		\end{subfigure}
		\quad
		\begin{subfigure}[b]{0.45 \linewidth}
			\centering
			\includegraphics[height=1 in]{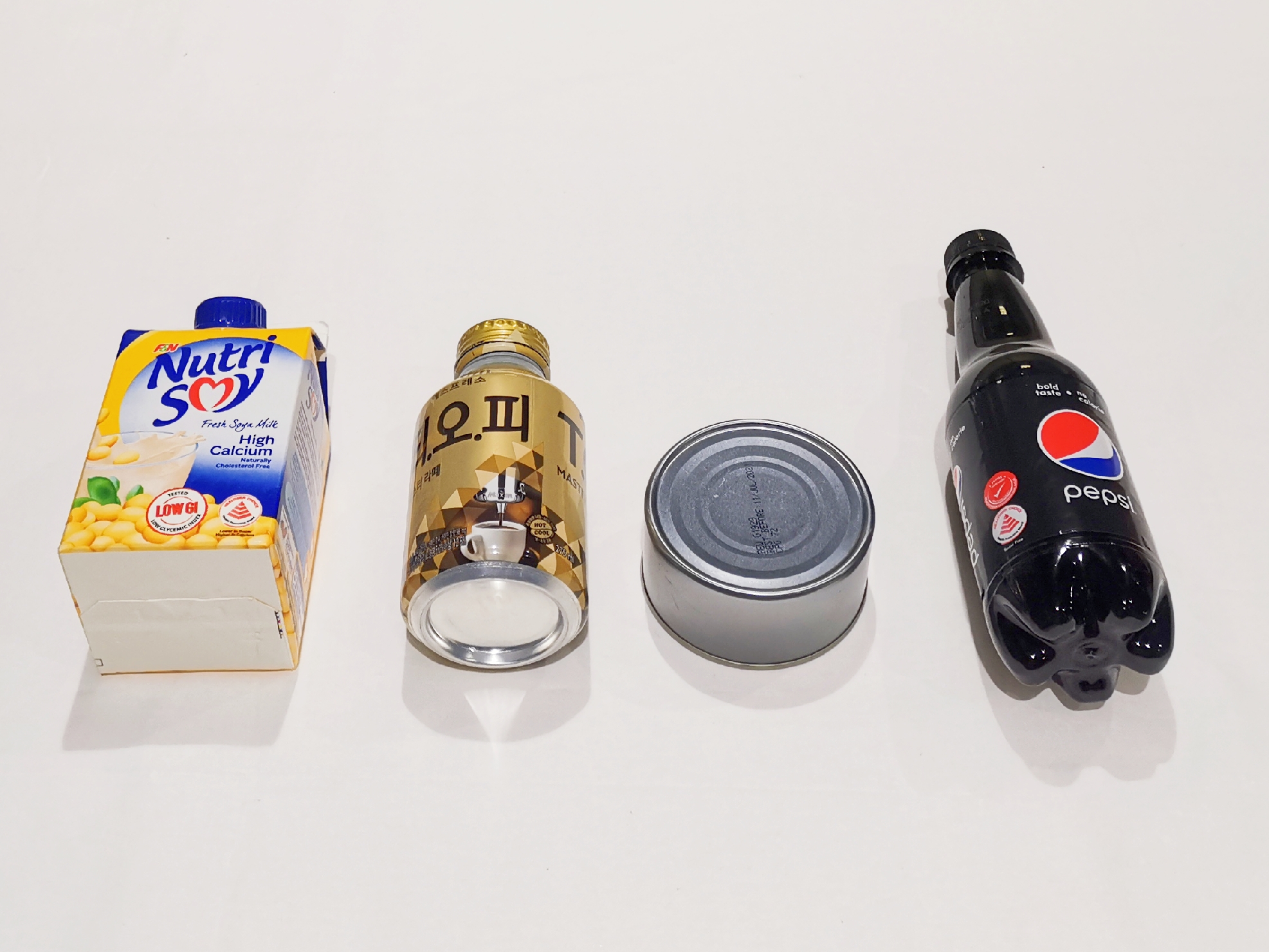} 
			\caption{EvTouch-Containers}
		\end{subfigure}
		\caption{(a) EvTouch-Objects dataset contains 36 object classes, including 20 objects from the YCB benchmarks (the first four rows). (b)  EvTouch-Containers contains four types of containers: coffee can, plastic soda bottle, soy milk carton, and metal tuna can. Each container is filled with five amount of the same liquid, resulting in 20 object classes.}
		\label{fig:datasets}
	\end{figure}

	We compared the methods using the recently developed event-based tactile datasets~\cite{taunyazov20event}. In brief, the datasets were collected using a 7-DoF Franka Emika Panda arm equipped with a Robotiq 2F-140 gripper, equipped with a NeuTouch event-based tactile sensor~\cite{taunyazov20event} and an ACES decoder~\cite{lee2019neuro} to decode the sensor signals into spikes. The Panda picked up a variety of different household objects to generate the two datasets:
	\begin{itemize}
		\item \textbf{EvTouch-Objects}: This dataset comprises tactile data from 36 object classes (Fig. \ref{fig:datasets}(a)). Among these objects, 26 are objects from the YCB dataset~\cite{calli2015benchmarking}, and the remaining 10 objects are deformable objects chosen to supplement the relatively rigid YCB objects. 
		To collect tactile data, the robot gripper grasped the object, and  lifted it off the table by 20 cm before placing it back onto the table.
		We used the data collected during the time from lifting an object to releasing it ($\approx$ 5 seconds). 
		For each object class, 20 samples were collected, yielding a total of 720 samples.
		\item \textbf{EvTouch-Containers}: This dataset includes tactile data for  four containers: a coffee can, a plastic soda bottle, a soymilk carton, and a metal tuna can (Fig. \ref{fig:datasets} (b)). These containers have a maximum volume of 250g, 400g, 300g, and 140g, respectively. Each container was filled with \{0\%, 25\%, 50\%, 75\%, 100\%\} of the respective maximum amount of water (or rice for   the open tuna can), resulting in 20 object classes. During the data collection, the robot gripper grasped each container, and then lifted it off the table by 5 cm. 
		We used the data collected during the time grasping an object to lifting and holding it for a while ($\approx$ 6.5 seconds in total). 
		There are a total of 300 samples (15 samples per object class). This dataset may be particularly challenging for tactile sensing since the  weights may not be easily distinguishable.
	\end{itemize}
	For both datasets, we used a bin duration of 0.02 seconds. Interested readers can find more details about the datasets in \cite{taunyazov20event} and the corresponding website\footnote{\url{https://clear-nus.github.io/visuotactile/}}.
	
	\subsection{Compared Methods}
	We compared the proposed TactileSGNet with three baseline spiking architectures (described below). All methods used LIF neurons and were implemented in PyTorch using a LIF-based framework~\cite{wu2019direct}.
	For fair comparison, all the methods share a similar network structure (number of layers, and number of units for respective layers) and similar hyperparameters. More precisely, the general network structure was {\small\texttt{Input-TAGConv-FC1(128)-FC2(256)-Voting}}, and we substituted the {\small\texttt{TAGConv}} layer with one of the following baselines:
	\begin{itemize}
		\item \textbf{MLP}, which uses a standard fully connected layer with 64 neurons to replace the TAGConv. This baseline represents the setup where minimal prior structure is introduced;
		\item \textbf{Grid-based CNN} where the tactile data was organized in a grid structure according to the spatial distribution of taxels~\cite{zapata2018non}. We set the size of each grid cell to 1 mm $\times$ 1 mm, yielding a grid of 13$\times$19 cells. The raw readings from each taxel were assigned to a grid cell and the remaining unfilled grid cells were zero-filled. 
		\item \textbf{GCN}, where we replaced TAGConv with the GCN~\cite{kipf2016semi}. As such, this baseline is similar to the state-of-the-art  TactileGCN~\cite{garcia2019tactilegcn}), except that the network is a SNN. 
	\end{itemize}
	Source code for our models is available at 
	\href{https://github.com/clear-nus/TactileSGNet}{\small\texttt{https://github.com/clear-nus/TactileSGNet}}.
	
	\subsection{Training and Evaluation Methodology}
	The parameters used in our models (and for training) are given in Table \ref{tab:hayperparameter_setting}. We split the data into a training set (80\%) and a test set (20\%) with equal class distribution. We optimized each model on the training dataset for 100 epochs using the Adam optimizer. Our comparison measure was accuracy on the test dataset. We repeated the training and test procedure for 10 rounds (with different initialization). 
	
\begin{figure}
		\centering
		\includegraphics[width=0.9\linewidth]{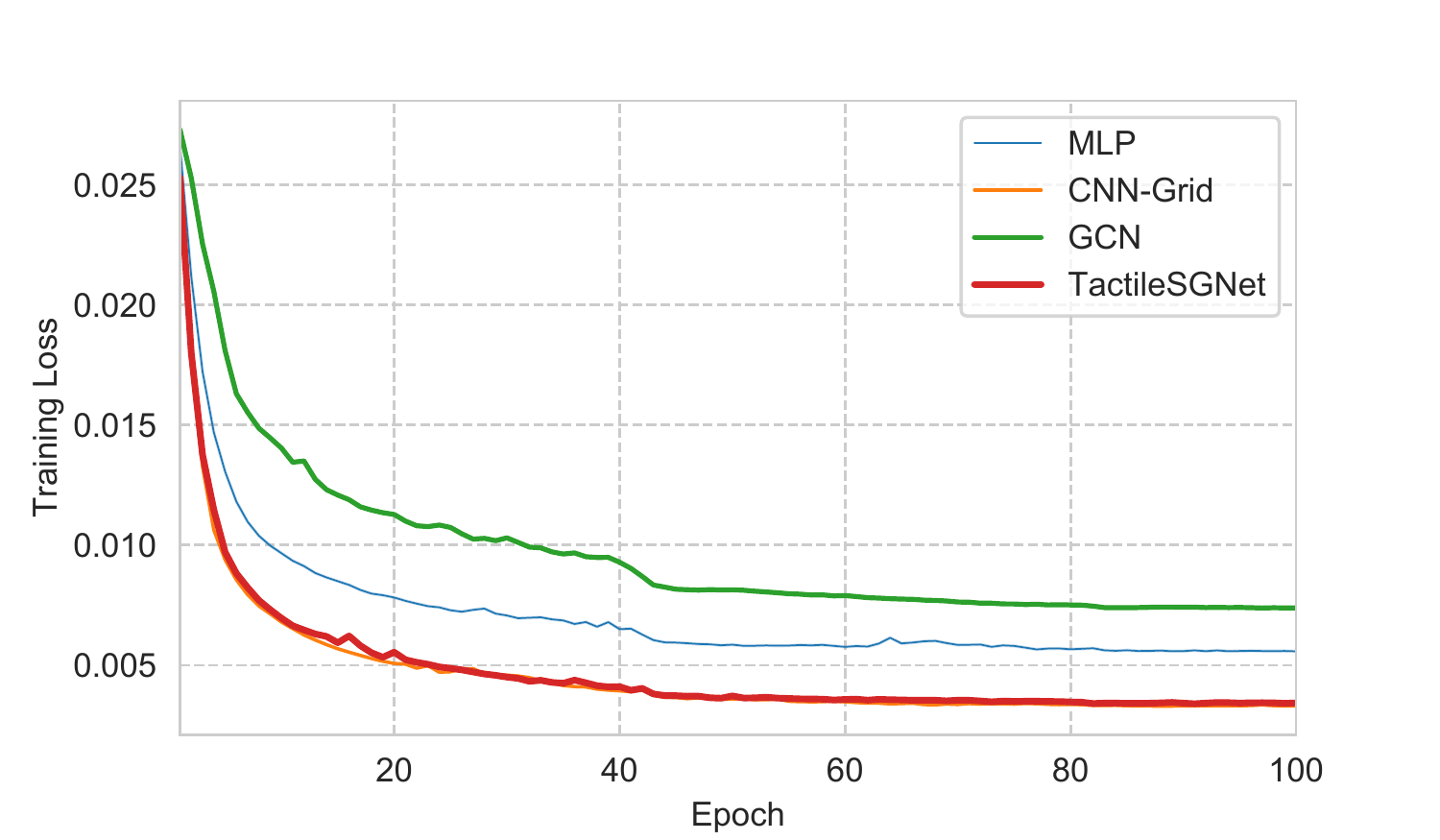}
		\includegraphics[width= 0.9 \linewidth]{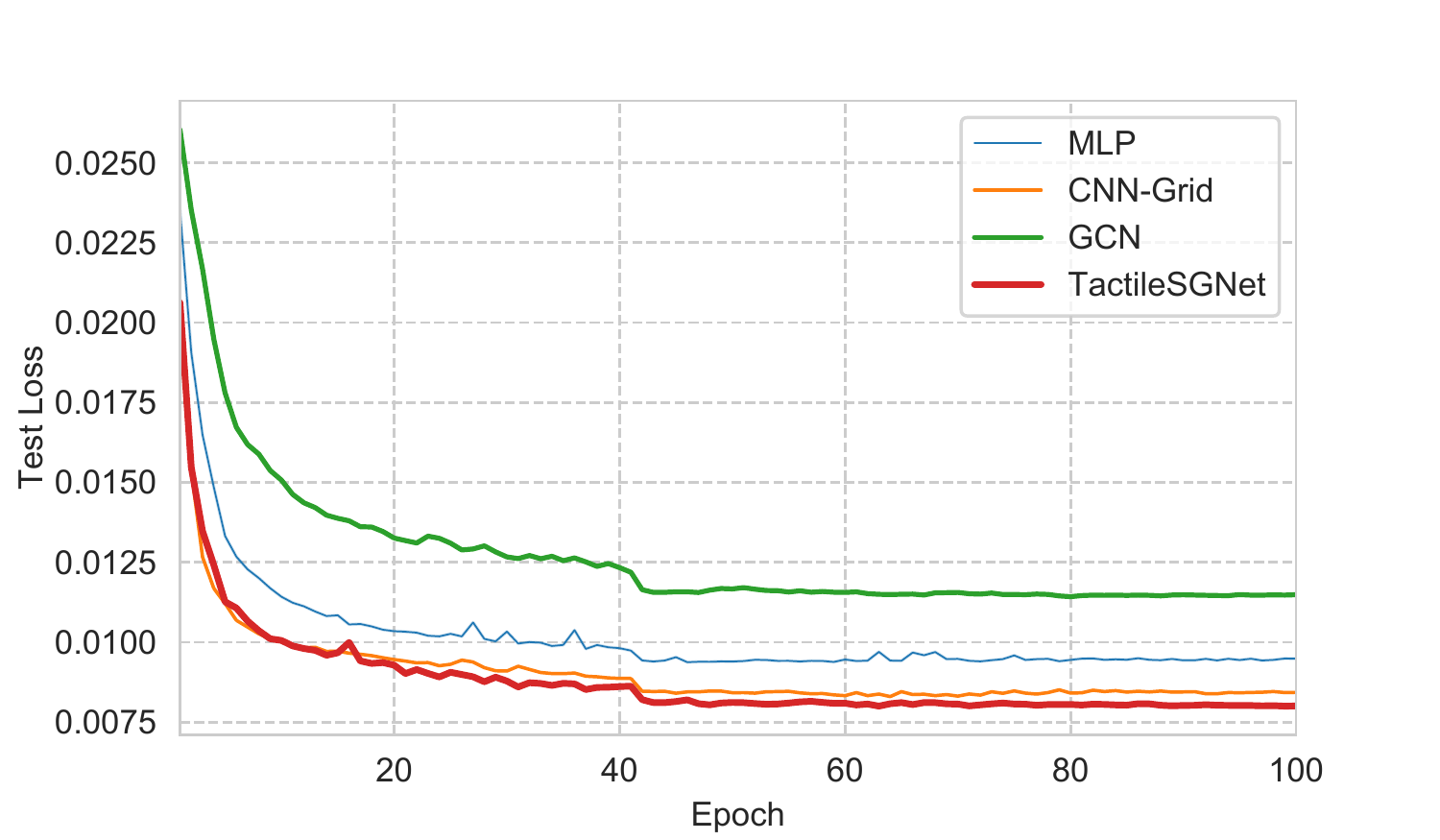} 
		\caption{Training and Test Losses as training progressed on EvTouch-Objects.}
		\label{fig:training_loss}
	\end{figure}

	We manually verified that all models were sufficiently trained by examining their training (testing) loss profiles. Figures \ref{fig:training_loss} show the training (test) losses as the iterations progressed in a representative run; the training loss and test loss for all the methods decreased quickly at the beginning epochs and then gradually converged. The TactileSGNet converged faster than the other methods and has both a lower training loss and test loss. 
	
	\begin{table}
		\centering
		\caption{\label{tab:hayperparameter_setting} Hyperparameter setting}
		\begin{tabular}{lc}
			\toprule[0.5pt]
			\bf{Parameter} & \bf{Value}   \\ 
			\midrule[0.5pt]
			Number of Layers &  3    \\
			Membrane potential threshold $u_T$ & 0.5 \\
			Potential reset parameter $u_R$  & 0\\
			Batch size & 1\\
			Decay factor of membrane potential  $\beta$ & 0.2 \\
			Learning rate & $1\times 10^{-3}$\\
			Width parameter of the approximated derivative $a$ & 0.5 \\
			\bottomrule[0.5pt]
		\end{tabular}
	\end{table}
		
	\subsection{Object Classification Performance}
	
	Table \ref{tab:acc_36} shows the mean accuracy with standard deviation over the 10 rounds. We observe that the TactileSGNet outperforms the other methods, with a mean accuracy of 89.44\% (on EvTouch-Objects dataset) and 64.17\% (on EvTouch-Containers), respectively. Among the baselines, the Grid-based CNN outperformed MLP and GCN the on both datasets. Compared to Grid-based CNN, our method has an accuracy improvement of about 1\% and 4\% on EvTouch-Objects dataset and EvTouch-Containers dataset, respectively. The reason for this difference may be the TAGConv layer was better able to process the graph-based representation which encoded the placement of the taxels. 
	
	\begin{table}
		\centering
		\caption{\label{tab:recognition_acc_dataset36} Test Accuracy of the Compared Methods on EvTouch-Objects  and EvTouch-Containers.}
		\begin{tabular}{lcc}
			\toprule[0.5pt]
			\bf{Method} & \bf{EvTouch-Objects} & \bf{EvTouch-Containers} \\ 
			\midrule[0.5pt]
			Grid-based CNN &  88.40 (1.14) & 60.17 (2.78)  \\ 
			MLP &   85.97 (0.85) & 58.83 (2.49) \\ 
			GCN & 85.14 (1.51) & 58.83 (2.84) \\
			\bf{TactileSGNet} & \bf{89.44 (0.55)} & \bf{64.17 (2.75)} \\ %\hline
			\bottomrule[0.5pt]
		\end{tabular}
		\label{tab:acc_36}
	\end{table}
	
	\begin{figure}
		\centering
		\includegraphics[width= 1 \linewidth]{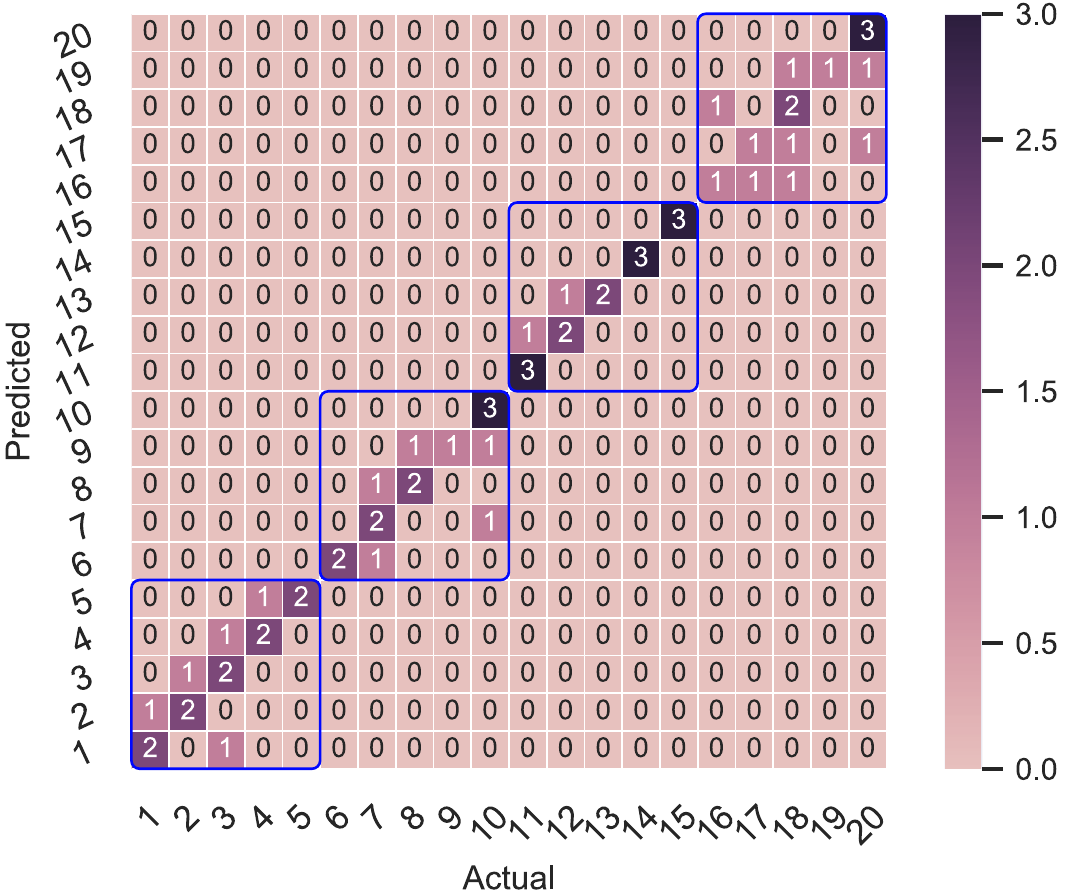} 
		\caption{TactileSGNet Confusion Matrix (on EvTouch-Containers dataset). Each blue rectangle denotes that the object classes within the rectangle are from the same container. Class 1-5 is for plastic soda bottle, class 6-10 is for tuna fish can, class 11-15 is for soy milk carton, and class 16-20 is for coffee can.}
		\label{fig:confusion_matrix}
	\end{figure}
	
	Fig \ref{fig:confusion_matrix} shows the confusion matrix for test set generated from one round on EvTouch-Containers (an illustrative sample). Surprisingly, we found that TactileSGNet was able to perfectly distinguish the different containers (blue rectangles); we had initially expected the model to confuse objects of similar rigidity such as the coffee and tuna cans. TactileSGNet was also able to recognize the container fullness with a relatively high accuracy. The different weight classes in soy milk carton were easier to classify --- possibly due to the softness of the container which allowed the sensor and model to detect the pressure exerted --- while the coffee can was more difficult (possibly due to container rigidity). We also see that many of the incorrect classifications are reasonable and among similar weights.

	\subsection{Impact of Graph Connectivity}
	
	In this section, we compare the three methods introduced in Section \ref{sec:tactile_graph} as input to TactileSGNet. For kNN, the value of k we considered ranged from 1 to 8. For the MST + $\sigma_d$ method, the distance threshold was set in the range of (0, 1.5, 2, 2.5, 3, 3.5, 4); note that the shortest distance between taxels on the NeuTouch is about 1.5 mm.
	
	Table \ref{tab:graph_connectivity} shows the accuracy of the TactileSGNet using different tactile graphs. 
	For the kNN method, the best mean accuracy is achieved when $k=7$ on EvTouch-Objects, and when $k=4$ on  EvTouch-Containers. For the MST + $\sigma_d$ method, the best mean accuracy is when $\sigma_d = 3$ on  EvTouch-Objects, and when $\sigma_d=1.5$ on  EvTouch-Containers. Overall, the best accuracy achieved by both kNN and MST + $\sigma_d$ methods is slightly higher on EvTouch-Objects and about 3\% higher on EvTouch-Containers than the manual method. 
	Tactile graphs with more edges (a larger $k$ or $\sigma_d$) did not necessarily result in a better accuracy. The proposed method appears robust to the number of edge connections in tactile graphs, and the deviation in the classification accuracy achieved with different connections was relatively small (below 1\% on EvTouch-Objects and 4\% on EvTouch-Containers).

	\begin{table}
		\centering
		\caption{\label{tab:graph_connectivity} Accuracy of TactileSGNet using different tactile graphs on the two datasets. $\langle k \rangle$ denotes the average node degree.}
		\resizebox{\linewidth}{!}{
			\begin{tabular}{ccccccc}
				\toprule[0.5pt]
				& Parameter &  $ \langle k \rangle$ & EvTouch-Objects & EvTouch-Containers\\ 
				\midrule[0.5pt]
				Manual & - & - & 89.44 (0.55) & 64.17 (2.75) \\ 	\hline
				\multirow{6}{*}{kNN} %undirected
				& k = 1 &  1  &88.75 (0.64) & 62.67 (2.53)\\
				& k = 2 &   2  &88.75 (0.79) & 63.33 (1.18) \\
				& k = 3 &   3  &  89.24 (1.00) &  66.00 (0.91) \\
				& k = 4 &    4  & 88.96 (0.69) & \bf{67.00 (1.39)} \\
				& k = 5 &    5  & 89.44 (0.44) & 65.67 (2.79)\\
				& k = 6 &   6  & 89.31 (0.67) &  62.00 (0.75) \\
				& k = 7 &  7 & \bf{89.65 (0.83)} & 64.67 (4.31)\\
				& k = 8 &  8 & 89.51 (0.83) &  60.33 (1.39)\\
				\hline
				\multirow{7}{*}{MST + $\sigma_d$ } 
				& $\sigma_d$ = 0.0 & $ 1.9 $ &  89.03 (0.64) & 63.67 (2.17)\\
				& $\sigma_d$ = 1.5 & $ 2.1 $ & 88.68 (0.93) & \bf{65.33 (3.80)} \\
				& $\sigma_d$ = 2.0 & $ 3.3 $ & 89.44 (0.55) & 63.00 (0.75)\\
				& $\sigma_d$ = 2.5 & $4.1$ & 89.31 (0.94) & 65.00 (2.36)\\
				& $\sigma_d$ = 3.0 &  $ 5.1 $ & \bf{89.51 (0.69)} & 64.33 (1.49) \\
				& $\sigma_d$ = 3.5 & $ 8.4 $ & 89.17 (0.94) &  63.00 (3.80) \\
				& $\sigma_d$ = 4.0 & $ 10.4 $ & 89.03 (0.85) & 62.00 (1.39)\\				
				\bottomrule[0.5pt]
			\end{tabular}
		}
	\end{table}

	\section{Conclusion}
	
	In this paper, we present a novel spiking graph neural network for event-based tactile learning. Compared to existing works, our method can exploit local topological structure of the taxels via an input graph. Experiments show that our method achieves higher performance compared to existing methods; using only event-based tactile data, TactileSGNet are able to distinguish various household objects with almost 90\% accuracy. More broadly, our results indicate that event-driven tactile perception can be effective and we hope that our findings will spur research in this promising area.

	\section*{Acknowledgments}
	This work was supported by the SERC, A*STAR, Singapore, through the National Robotics Program under Grant No. 172 25 00063.

	%%%%%%%%%%%%%%%%%%%%%%%%%%%%%%%%%%%%%%%%%%%%%%%%%%%%%%%%%%%%%%%%%%%%%%%%%%%%%%%%

	%%%%%%%%%%%%%%%%%%%%%%%%%%%%%%%%%%%%%%%%%%%%%%%%%%%%%%%%%%%%%%%%%%%%%%%%%%%%%%%%

	%%%%%%%%%%%%%%%%%%%%%%%%%%%%%%%%%%%%%%%%%%%%%%%%%%%%%%%%%%%%%%%%%%%%%%%%%%%%%%%%
	%\section*{APPENDIX}
	
	
	%\section*{ACKNOWLEDGMENT}

	%%%%%%%%%%%%%%%%%%%%%%%%%%%%%%%%%%%%%%%%%%%%%%%%%%%%%%%%%%%%%%%%%%%%%%%%%%%%%%%%
	
	%References are important to the reader; therefore, each citation must be complete and correct. If at all possible, references should be commonly available publications.
	\bibliographystyle{IEEEtran}
	\balance
	\bibliography{references}

\end{document}